\documentclass[11pt]{article}
\usepackage{epsfig}
\usepackage{amssymb}
\newcommand{\sect}[1]{ \section{#1} \setcounter{equation}{0} }

\newcommand{\xslash}{x \! \! \! /} 
\newcommand{\partialslash}{\partial \! \! \! /} 
\newcommand{\half}{\mbox{\small{$\frac{1}{2}$}}} 
 
\newcommand{\quarter}{\mbox{\small{$\frac{1}{4}$}}} 
\newcommand{\MSbar}{\overline{\mbox{MS}}} 
 
\newcommand{\Nf}{N_{\!f}}
\newcommand{\Nc}{N_{\!c}}
\newcommand{\NA}{N_{\!A}}

\begin{document}
\title{Critical exponent $\omega$ in the Gross-Neveu-Yukawa model at $O(1/N)$}
\author{J.A. Gracey, \\ Theoretical Physics Division, \\ 
Department of Mathematical Sciences, \\ University of Liverpool, \\ P.O. Box 
147, \\ Liverpool, \\ L69 3BX, \\ United Kingdom.} 
\date{} 
\maketitle 

\vspace{5cm} 
\noindent 
{\bf Abstract.} The critcal exponent $\omega$ is evaluated at $O(1/N)$ in
$d$-dimensions in the Gross-Neveu model using the large $N$ critical point 
formalism. It is shown to be in agreement with the recently determined three 
loop $\beta$-functions of the Gross-Neveu-Yukawa model in four dimensions. The
same exponent is computed for the chiral Gross-Neveu and non-abelian 
Nambu-Jona-Lasinio universality classes.  

\vspace{-16.5cm}
\hspace{13cm}
{\bf LTH 1138}

\newpage

\sect{Introduction.}

The Gross-Neveu (GN) model is a remarkably simple quantum field theory of a 
fermion field with a quartic interaction but which has a rich spectrum of 
properties with applications to many areas of physics. It was first studied at 
length in the particle physics context in \cite{1} but was introduced in 
\cite{2} and is also known as the Ashkin-Teller model. In \cite{1} it was shown 
that the theory was renormalizable in two dimensions and is asymptotically 
free. When the model is endowed with an $O(N)$ or $SU(N)$ symmetry it can be 
solved within the large $N$ expansion, \cite{1}, where it is apparent that 
there is dynamical symmetry breaking. The originally massless fermion field 
becomes massive in the true non-perturbative vacuum and contains a massive
$2$-fermion bound state particle in the spectrum. In many ways the Gross-Neveu 
model features a large cross-section of properties of some four dimensional 
gauge theories such as Quantum Chromodynamics (QCD) which means it is a simpler
forum for testing out ideas in higher dimensional quantum field theories. 
Unlike gauge theories the exact $S$-matrix can be written down for the 
Gross-Neveu model, \cite{3}, from which the full spectrum of bound states can 
be deduced. Also the exact mass gap which fixes the scale of the dynamically 
generated mass has been computed from the exact $S$-matrix, \cite{4}. In terms 
of other connections to physics, the Gross-Neveu model critical exponents 
derived from the renormalization group functions at the $d$-dimensional 
Wilson-Fisher fixed point for a particular value of $N$ have recently been 
shown to describe the critical dynamics of a phase transition in graphene, 
\cite{5}. This rather novel and perhaps surprising connection may offer a hope 
that the Gross-Neveu model, which is sometimes used as a theoretical laboratory
for a four dimensional gauge theory, could in fact become an experimental one. 
For instance, it has been suggested in \cite{5} that the transition in graphene
could be a paradigm of Standard Model spontaneous symmetry breaking. 

There is also perhaps more immediate interest in the Gross-Neveu model due to
recent connections with other four dimensional theories as well as AdS/CFT 
ideas, \cite{6,7,8}. For instance, the model is one of a set of theories which 
resides in the infinite tower of quantum field theories at the $d$-dimensional 
Wilson-Fisher fixed point. Another constituent of this universality class is 
the Gross-Neveu-Yukawa theory (GNY) which is renormalizable in four dimensions 
with the connection to the Gross-Neveu model being illuminated in \cite{9}. For
instance, both have the same core boson-fermion interaction differing only in 
the purely bosonic sector which is present to ensure renormalizability in their
respective critical dimensions. The current term for such a connection is 
ultraviolet completion which has opened up interesting ideas for aspects of the
Standard Model. For instance, a $4$-fermi interaction is non-renormalizable in 
four dimensions but by contrast the Gross-Neveu-Yukawa theory is 
renormalizable. The former was used as an effective theory in the early years 
of trying to understand weak interactions prior to the discovery of mass 
generation via spontaneous symmetry breaking and the Higgs mechanism. Within 
the Gross-Neveu-Yukawa model there is a bosonic field which is parallel to the 
Higgs field in that it also has a quartic self-interaction. More recently it 
has been shown that for a low value of $N$ when there is an $SU(N)$ symmetry 
the theory has an emergent supersymmetry, \cite{7}. By this it is meant that 
when the fixed point spectrum is determined from the zeros of the two 
$\beta$-functions, there is one solution where the coupling constants are equal
and moreover the anomalous dimensions of the fields evaluated at that fixed 
point are equal. This offers an interesting possibility of exploring a new way 
of going beyond the Standard Model. As such the Gross-Neveu-Yukawa model is an 
excellent forum to explore such new ideas. Necessary to achieve this at high 
precision is knowledge of the renormalization group functions at high loop 
order. Recently the three loop renormalization group functions of the
Gross-Neveu-Yukawa model were computed in the $\MSbar$ (modified minimal 
subtraction) scheme, \cite{10}, which extended the two loop results of 
\cite{11}. Given recent advances in multiloop techniques it would be reasonable
to expect four loop results in the near future. Aside from the graphene 
connection through the Gross-Neveu universality class, such high order loop 
computations are required if one is to extend the Standard Model 
renormalization group functions to the same precision.  

One aspect which naturally arises in this context is the need to have checks on
such intensely complicated computations at four loops. This is provided for via
the critical point large $N$ expansion which was developed originally for the 
$O(N)$ $\phi^4$ theory or equivalently the Ising model universality class in 
\cite{12,13,14}. There the first three terms of the $d$-dimensional large $N$
critical exponents were determined and these contain information on the 
renormalization group functions of all the theories lying in the same 
universality class. In particular expanding the exponents in powers of 
$\epsilon$ where the spacetime dimension $d$ is set to 
$d$~$=$~$D$~$-$~$2\epsilon$ where $D$ is the critical dimension of a theory, 
there is a one-to-one correspondence with the $\epsilon$ expansion of the 
renormalization group functions at that theory's Wilson-Fisher fixed point. So 
the $d$-dimensional large $N$ critical exponents contain information on parts 
of the renormalization group functions which have not yet been computed 
explicitly. If several orders in $1/N$ are available such information is a 
useful independent blind test of any new loop order evaluation. The method used 
to determine the large $N$ critical exponents of the Ising model universal 
theory has been extended to the Gross-Neveu universality class as well as 
related classes involving a simple core boson-fermion Yukawa type interaction, 
\cite{15,16,17,18,19,20,21}. These exponents have proved useful in comparing 
with explicit Gross-Neveu perturbative computations. However, in order to 
extract {\em new} information for the four dimensional Gross-Neveu-Yukawa 
theory from these results beyond what is currently available in perturbation
theory one key large $N$ critical exponent has yet to be determined which is 
denoted by $\omega$. It relates to the $\beta$-function slope at criticality 
relative to the four dimensional theory and it is the purpose of this article 
to compute it at leading order in large $N$. This is not only for the 
Gross-Neveu universality class, which contains the Gross-Neveu-Yukawa theory, 
but also for the chiral Gross-Neveu and non-abelian Nambu-Jona-Lasinio 
universality classes. The reason why these other classes are considered is that
they are endowed with a combination of symmetries different from the 
Gross-Neveu class such as continuous chiral symmetry with fermions lying in 
multiplets of non-abelian groups. As such they are closer to the symmetries of 
the Standard Model. Aside from this connection to Standard Model 
renormalization group functions knowledge of $\omega$ will add to the 
information which is accumulating on our understanding of ultraviolet 
completeness and $d$-dimensional conformal field theories.

The article is organized as follows. The large $N$ critical point formalism of
\cite{12,13} extended to the Gross-Neveu model in \cite{15} is reviewed in 
section $2$ in the context of evaluating $\omega$ at $O(1/N)$. The explicit
equations from which $\omega$ is determined at this order are constructed in
section $3$ where an underlying two loop large $N$ graph which is required is
evaluated using massless coordinate space techniques. The main result for
$\omega$ at $O(1/N)$ is given here and its $\epsilon$ expansion reconciled with
the four dimensional renormalization group functions. The formalism of these
two sections is extended in the subsequent section to the other two $4$-fermi
universality classes and $\omega$ is deduced for each. Finally, we provide
conclusions in section $5$.

\sect{Formalism.}

We begin by recalling the relevant aspects of the Gross-Neveu and 
Gross-Neveu-Yukawa models and the large $N$ formalism in order to be able to
compute the critical exponent $\omega$ at $O(1/N)$. First, the respective
Lagrangians are, \cite{1}, 
\begin{equation}
L_{\mbox{\footnotesize{GN}}} ~=~ i \bar{\psi}^i \partialslash \psi^i ~+~
\frac{1}{2} g \sigma \bar{\psi}^i \psi^i ~-~ \frac{1}{2} \sigma^2
\label{laggn}
\end{equation}
for the two dimensional Gross-Neveu (GN) model where $1$~$\leq$~$i$~$\leq$ $N$, 
and, \cite{9},
\begin{equation}
L_{\mbox{\footnotesize{GNY}}} ~=~ i \bar{\psi}^i \partialslash \psi^i ~+~
\frac{1}{2} \partial_\mu \sigma \partial^\mu \sigma ~+~
\frac{1}{2} g_1 \sigma \bar{\psi}^i \psi^i ~+~ \frac{1}{24} g_2^2 \sigma^4
\label{laggny}
\end{equation}
for the four dimensional Gross-Neveu-Yukawa (GNY) model. In both cases the
fermion field $\psi^i$ has an $SU(N)$ symmetry with $1$~$\leq$~$i$~$\leq$~$N$
and $\sigma$ is a bosonic field. In (\ref{laggn}) this field is actually an
auxiliary and it can be eliminated to produce the Lagrangian, \cite{1}, 
\begin{equation}
L_{\mbox{\footnotesize{GN}}} ~=~ i \bar{\psi}^{i} \partialslash \psi^{i} ~+~ 
\frac{1}{2} g^2 ( \bar{\psi}^i \psi^i )^2
\label{laggn4fermi}
\end{equation}
which is asymptotically free and renormalizable in two dimensions, \cite{1}. By
contrast in (\ref{laggny}) $\sigma$ has a canonical kinetic term and hence it 
cannot be eliminated from the Lagrangian which is renormalizable in four 
dimensions. Both Lagrangians share common properties. The coupling constants of 
each theory are dimensionless in the respective critical dimensions which means
that the canonical dimension of $\sigma$ is the same in both Lagrangians. This 
is not an accident since the Gross-Neveu-Yukawa model is the ultraviolet 
completion of (\ref{laggn}) and both are contained within the same universality
class at their respective Wilson-Fisher fixed points. In addition both 
(\ref{laggn}) and (\ref{laggny}) have one interaction the same which we will 
term the core interaction and which is the fundamental interaction of the 
underlying universality class at the respective Wilson-Fisher fixed points. 
While the canonical dimension of the fermion is different in the critical 
dimensions of each Lagrangian it can be expressed as $\half(d-1)$ where $d$ is 
the spacetime dimension. Thus the $4$-fermi interaction of (\ref{laggn4fermi}) 
would be a dimension six operator if it was present in (\ref{laggny}) and hence
would correspond to a non-renormalizable operator. In noting that both theories
have a connection through the ultraviolet completion it has been known for a 
long time that both can be studied via the large $N$ expansion where $1/N$ is 
regarded as a small perturbative parameter, \cite{1,9}. As $1/N$ is a 
dimensionless parameter it turns out that one can obtain information about the 
underlying universality class which contains both (\ref{laggn}) and 
(\ref{laggny}). In particular it is possible to compute the critical exponents 
of the universal theory order by order in $1/N$ in an arbitrary spacetime 
dimension $d$ and moreover to {\em three} orders in $1/N$. This was first 
carried out for the $O(N)$ nonlinear $\sigma$ model and $\phi^4$ theory in 
\cite{12,13,14}. Thereafter the same information was determined for the $O(N)$ 
Gross-Neveu universality class in \cite{15,16,17,18,19,20,21}. These critical 
exponents contain all the information on the renormalization group functions in
the tower of theories lying along a Wilson-Fisher fixed point in 
$d$-dimensions. For instance, expanding the exponents in 
$d$~$=$~$2$~$-$~$2\epsilon$ or $d$~$=$~$4$~$-$~$2\epsilon$ one obtains the 
$\epsilon$ expansion of the critical exponents derived from the respective 
renormalization group functions of (\ref{laggn}) and (\ref{laggny}). For the
latter case we now focus on computing $\omega$ at $O(1/N)$ in order to obtain 
information on the Gross-Neveu-Yukawa renormalization group functions. 

The first stage is to write down the universal Lagrangian used for the large 
$N$ formalism which is 
\begin{equation}
L_{\mbox{\footnotesize{GN}}} ~=~ i \bar{\psi}^i \partialslash \psi^i ~+~
\frac{1}{2} \tilde{\sigma} \bar{\psi}^i \psi^i ~-~ 
\frac{1}{2g^2} \tilde{\sigma}^2 ~.
\label{laggnN}
\end{equation}
The only difference between (\ref{laggn}) and (\ref{laggnN}) is that the 
coupling constant has been rescaled out of the core interaction which will be 
the only interaction relevant in the universal theory. In (\ref{laggny}), for 
instance, the quartic $\sigma$ interaction, which is akin to the Higgs 
interaction in the Standard Model, is included with the core interaction to 
ensure the field theory is renormalizable in the critical dimension of $4$. By 
contrast in the universal theory (\ref{laggnN}) such an interaction emerges 
naturally through box graphs with four external $\sigma$ legs and only the 
$\sigma \bar{\psi}^i \psi^i$ interaction. Such a property was first noted in 
\cite{22} for the critical point equivalence of the non-abelian Thirring model 
and QCD in the large $\Nf$ expansion where $\Nf$ is the number of (massless) 
quark flavours as opposed to an expansion in $1/\Nc$ where $\Nc$ is the number 
of colours. The final term of (\ref{laggnN}) is only included to make the 
connection with (\ref{laggn}) and is not central to setting up the $1/N$ 
formalism developed in \cite{12,13}. It is actually the first two terms of 
(\ref{laggnN}) which define the scaling behaviour of the fields at criticality.
From these we can define the respective full dimensions of the $\psi^i$ and 
$\sigma$ fields as, \cite{15}, 
\begin{equation}
\alpha ~=~ \mu ~+~ \half \eta ~~~,~~~ \beta ~=~ 1 ~-~ \eta ~-~ \chi
\end{equation}
where $d$~$=$~$2\mu$ for shorthand, $\eta$ is the anomalous dimension exponent
of the $\psi^i$ field and $\chi$ is the exponent associated with the anomalous
dimension of the core interaction. Once these dimensions are specified then the
asymptotic scaling forms of the respective propagators in the approach to 
criticality can be specified as 
\begin{equation}
\psi (x) ~ \sim ~ \frac{A\xslash}{(x^2)^\alpha} ~~~,~~~
\sigma(x) ~ \sim ~ \frac{B}{(x^2)^\beta}
\label{asymprop}
\end{equation}
in coordinate space where we use the same letter as the field to denote the
propagator. The canonical part of $\alpha$ is not inconsistent with the
canonical dimension of $\psi^i$ due to the presence of $\xslash$ in the 
numerator of the scaling form of the propagator. Also $A$ and $B$ are 
$x$-independent amplitudes which always appear within computations in the 
combination $z$~$=$~$A^2B$. We will work in coordinate space throughout and use
the same notation as \cite{15}. We recall that the $1/N$ expansion of various 
quantities will be written as 
\begin{equation}
\eta ~=~ \sum_{n=1}^\infty \frac{\eta_n}{N^n} ~~~,~~~ 
z ~=~ \sum_{n=1}^\infty \frac{z_n}{N^n}
\label{expexp}
\end{equation}
for example. As these and other quantities have already been computed in the
Gross-Neveu universality class we recall the various $O(1/N)$ expressions here 
for completeness. We have, \cite{15,17,18},
\begin{equation}
\eta_1 ~=~ -~ \frac{2(\mu-1)\Gamma(2\mu-1)}
{\mu\Gamma^3(\mu)\Gamma(1-\mu)} ~~~,~~~ 
\chi_1 ~=~ \frac{\mu\eta_1}{(\mu-1)} ~~~,~~~ 
z_1 ~=~ \frac{1}{2} \Gamma^2(\mu) \eta_1 
\label{gnloexp}
\end{equation}
where we note that throughout we take the spinor trace convention as
$\mbox{tr} I$~$=$~$2$. While this is consistent when comparing with two
dimensional perturbation theory in order to compare any exponent with the 
critical exponents derived from four dimensional renormalization group 
functions one replaces $N$ by $2N$. This is because within Feynman graphs a
factor of $N$ is always associated with a closed fermion loop. The exponents of
(\ref{gnloexp}) were determined by solving the skeleton Schwinger-Dyson
equations for the two $2$-point functions order by order at criticality using 
the asymptotic scaling forms (\ref{asymprop}), \cite{15,17,18}.

The procedure to determine $\omega$ at $O(1/N)$ follows the same route as that
to find $\eta_1$ except that the leading asymptotic scaling forms of the
propagators are extended to 
\begin{equation}
\psi(x) ~\sim~ \frac{A\xslash}{(x^2)^\alpha} \left[ 1 + A^\prime(x^2)^\omega
\right] ~~~,~~~
\sigma(x) ~\sim~ \frac{B}{(x^2)^\beta} \left[ 1 + B^\prime(x^2)^\omega
\right] ~.
\label{asympropom}
\end{equation}
The additional terms represent corrections to scaling and as such depend on
the associated exponent which is $\omega$ which has the expansion
\begin{equation}
\omega ~=~ \sum_{n=0}^\infty \frac{\omega_n}{N^n}
\end{equation}
with $\omega_0$~$=$~$\mu$~$-$~$2$. It relates to the slope of the 
$\beta$-function at the critical point. In \cite{15,17,18} the exponent 
corresponding to the $\beta$-function of (\ref{laggn}) has the value of 
$(\mu-1)$ as the canonical dimension of that theory which has a critical 
dimension of $2$. The additional amplitudes $A^\prime$ and $B^\prime$ are also 
$x$-independent. While (\ref{asympropom}) corresponds to the critical behaviour 
of the propagators in the asymptotic limit to the fixed point the corresponding
expressions for the $2$-point functions are needed in order to solve the 
skeleton Schwinger-Dyson equations. Ordinarily one merely inverts the 
propagators to find these. However that is only valid in momentum space and we 
require the coordinate space forms given that we solve the Schwinger-Dyson 
equations in coordinate space. The process is the same but we have to map 
(\ref{asymprop}) to momentum space first using the Fourier transform, 
\cite{12,13},
\begin{equation}
\frac{1}{(x^2)^\alpha} ~=~ \frac{a(\alpha)}{2^{2\alpha}} \int_k
\frac{e^{ikx}}{(k^2)^{\mu-\alpha}}
\end{equation}
for any exponent $\alpha$ where
\begin{equation}
a(\alpha) ~=~ \frac{\Gamma(\mu-\alpha)}{\Gamma(\alpha)}
\end{equation}
for shorthand and $\int_k$~$=$~$\int \frac{d^dk}{(2\pi)^d}$. Then the 
propagator is inverted and the inverse Fourier transform used. Consequently the
asymptotic scaling forms of the full $2$-point functions with corrections to 
scaling are, \cite{15}, 
\begin{eqnarray}
\psi^{-1}(x) & \sim & \frac{r(\alpha-1)\xslash}{A(x^2)^{2\mu-\alpha+1}}
\left[ 1 - A^\prime s(\alpha-1)(x^2)^\omega \right] \\
\sigma^{-1}(x) & \sim &\frac{p(\beta)}{B(x^2)^{2\mu-\beta}}
\left[ 1 - B^\prime q(\beta) (x^2)^\omega \right]
\label{asymptwoom}
\end{eqnarray}
where the various functions are defined by
\begin{eqnarray}
p(\beta) &=& \frac{a(\beta-\mu)}{a(\beta)} ~~~,~~~
r(\alpha) ~=~ \frac{\alpha p(\alpha)}{(\mu-\alpha)} \nonumber \\
q(\beta) &=& \frac{a(\beta-\mu+\omega)a(\beta-\omega)}{a(\beta-\mu)a(\beta)}
~~~,~~~ s(\alpha) ~=~ \frac{\alpha(\alpha-\mu)q(\alpha)}{(\alpha-\mu+\omega)
(\alpha-\omega)} ~.
\end{eqnarray}
The first two functions are central to the determination of $\eta$ in large $N$
while the second pair will be used to find $\omega_1$. 

{\begin{figure}[hb]
\begin{center}
\includegraphics[width=11cm,height=3.5cm]{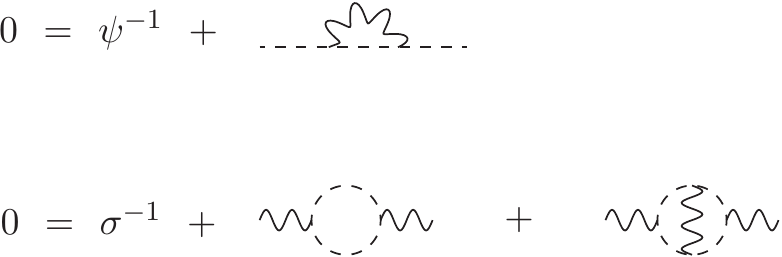}
\end{center}
%\vspace{0.5cm}
\caption{Leading order skeleton Schwinger-Dyson equations used to determine
$\omega_1$ in the Gross-Neveu-Yukawa model.}
\end{figure}}

\sect{Critical exponent $\omega_1$.}

The method to find $\omega_1$ is to substitute the asymptotic scaling forms for
the propagators and $2$-point functions into the skeleton Schwinger-Dyson
equations. The relevant Feynman diagrams are given in Figure $1$ which merit
several comments. First, the ordering of large $N$ graphs differs from that of
ordinary coupling constant perturbation theory in that the corresponding
parameter is $1/N$. However, as $N$ can emerge from the evaluation of a graph,
when, for example, there is a closed fermion loop, then two and three loop 
graphs can actually be the same order in large $N$ perturbation theory. This is
the case in scalar $O(N)$ theories but the same feature does not occur until 
$O(1/N^2)$ in the Gross-Neveu universality class. This is because a loop with a
trace over an odd number of $\gamma$-matrices vanishes. Also a reordering can 
occur for other reasons which we will discuss shortly. As the parameter $z$ 
also arises in the value of a graph and is $O(1/N)$ one counts a $\psi^i$ line 
as unity but a $\sigma$ field is $O(1/N)$. So in Figure $1$ the first graph of 
the $\sigma$ $2$-point is regarded as $O(N)$ but the subsequent graph is 
$O(1)$. There are no other graphs contributing to this $2$-point function at 
$O(1)$. In perturbation there would be additional graphs such as self-energy 
corrections to the leading order graph. These are not present in the large $N$ 
expansion since these contributions are contained in the anomalous parts of 
$\alpha$ and $\beta$. Including such self-energy corrected graphs would lead to 
overcounting, \cite{12}. When substituting (\ref{asympropom}) and 
(\ref{asymptwoom}) into the graphs of Figure $1$ the subsequent algebraic 
equations for each $2$-point function decouple into two equations. This is 
because the corrections to scaling terms have a different power of $x^2$. So, 
for example, the set at leading order is
\begin{equation}
0 ~=~ r(\alpha - 1) ~+~ z ~~~,~~~ 
0 ~=~ p(\beta) ~+~ N z 
\label{loeta}
\end{equation} 
from which the earlier values for $\eta_1$ and $z_1$ follow. 

{\begin{figure}[ht]
\begin{center}
\includegraphics[width=6cm,height=2.7cm]{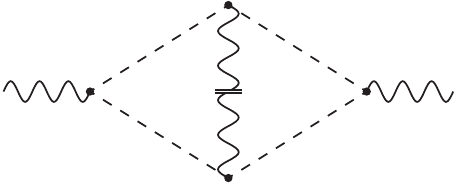}
\end{center}
%\vspace{0.5cm}
\caption{Diagram corresponding to $\Pi_{1B}$ where the double line indicates 
the inclusion of the correction to scaling exponent.}
\end{figure}}

The remaining set require a more careful treatment, \cite{17,18,20}, due to a 
reordering alluded to earlier. This arises from considering the value of the 
function $q(\beta)$ when the canonical dimensions for the $\sigma$ field and 
$\omega$ are substituted. In particular the function $a(\beta-\mu+\omega)$ is 
$O(1/N)$. In the scalar $O(N)$ theory considered in the original large $N$ 
treatment of \cite{12,13} the corresponding function was $O(1)$ for the 
$\beta$-function of the two dimensional theory. With respect to the four 
dimensional $\beta$-function of that universality class there is a similar 
reordering to that of the Gross-Neveu universality class, \cite{23}. The upshot
is that the two loop graph of the $\sigma$ skeleton Schwinger-Dyson equation 
cannot be omitted and the two equations required for $\omega_1$ are 
\begin{eqnarray}
0 &=& r(\alpha-1)[1-A^\prime s(\alpha-1)(x^2)^\omega] ~+~ z[1+(A^\prime
+ B^\prime)(x^2)^\omega] \nonumber \\
0 &=& p(\beta) [1-B^\prime q(\beta)(x^2)^\omega]
~+~ Nz [1+2A^\prime (x^2)^\omega] \nonumber \\
&& -~ \frac{1}{2} Nz^2 [ \Pi_1 + (\Pi_{1A}A^\prime
+ \Pi_{1B}B^\prime)(x^2)^\omega]
\label{loomega}
\end{eqnarray}
where we have split the contribution from the two loop $\sigma$ graph into its
contributions relative to scaling corrections. This is because one contribution
from the parts corresponding to the corrections to scaling are of the same 
order in $1/N$ as that from $q(\beta)$. The leading term, $\Pi_1$, is not 
needed for the equations determining $\eta_1$, (\ref{loeta}), but are necessary
for $\eta_2$ together with its counterpart in the $\psi^i$ $2$-point equation,
\cite{15}. The terms of (\ref{loomega}) which involve $(x^2)^\omega$ decouple 
from the corresponding terms of (\ref{loeta}) to leave two equations involving 
the correction amplitudes $A^\prime$ and $B^\prime$. Representing these as a 
$2$~$\times$~$2$ matrix the consistency equation which determines $\omega_1$ is
given by setting the determinant of this matrix to zero similar to the exponent
for the critical $\beta$-function slope for the two dimensional Gross-Neveu 
model, \cite{17,18,20}.

Examining this consistency equation the only quantity which is required for
finding $\omega_1$ is the value of the integral denoted by $\Pi_{1B}$. The
notation is such that the letter $B$ means that the correction to scaling is
included on the $\sigma$ line and the actual graph is illustrated in Figure
$2$. From the consistency equation the large $N$ reordering that requires
$\Pi_{1B}$ for $\omega_1$ is such that the other correction to scaling of this
two loop graph, $\Pi_{1A}$, will first arise in the determination of 
$\omega_2$. In Figure $2$ the double line on the $\sigma$ line indicates that
the exponent of that line is $\beta$~$-$~$\omega$ whereas the exponent on each
fermion line is $\alpha$. Unlike conventional perturbation theory the exponents 
of these lines depend on the parameter $N$ and moreover the coefficient of that
term in the exponent are the terms of the expansions such as (\ref{expexp}).
So the graph aside from being a function of $\mu$ due to the canonical
dimensions of the fields is also a function of $N$. However, for $\omega_1$
only the leading term in $1/N$ is required. The $O(1/N)$ correction to
$\Pi_{1B}$ will contribute to $\omega_2$. 

{\begin{figure}[hb]
\begin{center}
\includegraphics[width=5.5cm,height=3.5cm]{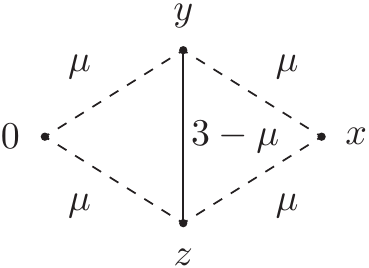}
\end{center}
%\vspace{0.5cm}
\caption{Feynman graph corresponding to $\Pi_{1B}$ with exponents included
where $x$, $y$ and $z$ are the location of the vertices in coordinate space.}
\end{figure}}

Substituting the canonical dimensions for the fields and $\omega$ produces the 
graph of Figure $3$. While the exponents of the lines are non-unit it is 
possible to evaluate the leading order term of the graph as a function of $\mu$ 
{\em exactly}. We have included the location of the internal and external 
vertices in Figure $3$ as we use coordinate space integration techniques 
developed in \cite{12,13}. In coordinate space the integration is over the 
location of the vertices. The first step of the evaluation is to apply a 
conformal transformation based on the left external vertex in the language of 
\cite{13} to the graph which produces the graph of Figure $4$. The 
transformation produces an overall minus sign as indicated there and there is
still a trace over the two remaining fermion lines. The resulting triangle with
scalar propagators is what is termed a one step from uniqueness triangle since 
the sum of the exponents comprising that triangle sum to $(\mu-1)$. For a
triangle to be unique the sum of exponents has to be $\mu$. The one step
from uniqueness rule has been given in \cite{24} and by applying it to this
integral the resulting set of graphs each involve simple coordinate space chain
integrals in the language of \cite{12,13}. This produces  
\begin{equation}
\Pi_{1B} ~=~ -~ \frac{2(\mu^2-4\mu+2)}{(\mu-1)^2(\mu-2)\Gamma^2(\mu)} ~+~
O \left( \frac{1}{N} \right) 
\label{valpi1b}
\end{equation}
at leading order. 

With this value for the integral it is a straightforward exercise to evaluate 
the determinant of the $\omega_1$ consistency equation at leading order in 
$1/N$ to obtain one of our main results 
\begin{equation}
\omega_1 ~=~ -~ \frac{2(2\mu-1)(\mu-2)}{(\mu-1)} \eta_1 ~.
\label{omexp}
\end{equation} 
From (\ref{omexp}) the $\epsilon$ expansion near $d$~$=$~$4$~$-$~$2\epsilon$
dimensions produces 
\begin{equation}
\omega_1 ~=~ 12 \epsilon^2 ~-~ 14 \epsilon^3 ~-~ 11 \epsilon^4 ~+~ 
\left[ 24 \zeta_3 - \frac{19}{2} \right] \epsilon^5 ~+~ O(\epsilon^6) ~.
\end{equation} 
Also in various odd dimensions we have 
\begin{eqnarray}
\left. \frac{}{} \omega \right|_{d=3} &=& -~ \frac{1}{2} ~+~ 
\frac{32}{3\pi^2 N} ~+~ O \left( \frac{1}{N^2} \right) \nonumber \\
\left. \frac{}{} \omega \right|_{d=5} &=& \frac{1}{2} ~+~ 
\frac{512}{15\pi^2 N} ~+~ O \left( \frac{1}{N^2} \right) 
\end{eqnarray}
with the $\epsilon$ expansion around $d$~$=$~$2$~$-$~$2\epsilon$ being 
non-singular.

{\begin{figure}[ht]
\begin{center}
\includegraphics[width=7cm,height=3.5cm]{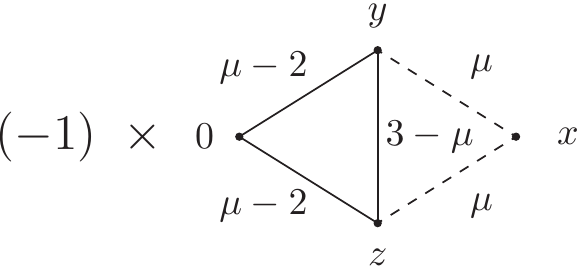}
\end{center}
%\vspace{0.5cm}
\caption{Feynman graph corresponding to $\Pi_{1B}$ after conformal left
transformation.}
\end{figure}}

The final stage of the exercise of computing $\omega_1$ is to check it against
the explicit two and three loop $\MSbar$ perturbative $\beta$-functions, 
\cite{10,11}. We recall the expressions for the $\beta$-functions of \cite{10}
in the notation of that paper are 
\begin{eqnarray}
\beta_1(g_1,g_2) &=& -~ 2 \epsilon g_1 ~+~ [2 N + 3] g_1^2 \nonumber \\
&& +~ 3 [ - 16 N g_1^2 - 3 g_1^2 - 64 g_1 g_2 + 64 g_2^2 ] \frac{g_1}{8} 
\nonumber \\ 
&& +~ [ 224 N^2 g_1^3 + 864 \zeta_3 N g_1^3 + 134 N g_1^3 
+ 912 \zeta_3 g_1^3 - 697 g_1^3 + 5760 N g_1^2 g_2 \nonumber \\
&& ~~~ + 8064 g_1^2 g_2 - 5760 N g_1 g_2^2 
+ 17472 g_1 g_2^2 - 13824 g_2^3 ] \frac{g_1}{64} \nonumber \\
&& +~ [ a_1 g_1^4 + a_2 g_1^3 g_2 + a_3 g_1^2 g_2^2 + a_4 g_1 g_2^3
+ a_5 g_2^4 ] g_1 N^3 ~+~ O(g_i^6) \nonumber \\
\beta_2(g_1,g_2) &=& -~ 2 \epsilon g_2 ~+~ 
[ - N g_1^2 + 4 N g_1 g_2 + 36 g_2^2 ] \nonumber \\ 
&& +~ [ 4 N g_1^3 + 7 N g_1^2 g_2 - 72 N g_1 g_2^2 - 816 g_2^3]
\nonumber \\ 
&& +~ [ - 628 N^2 g_1^4 - 384 \zeta_3 N g_1^4 + 5 N g_1^4 
+ 3472 N^2 g_1^3 g_2 - 3744 \zeta_3 N g_1^3 g_2 \nonumber \\
&& ~~~ - 8790 N g_1^3 g_2 
- 3456 N^2 g_1^2 g_2^2 + 31104 \zeta_3 N g_1^2 g_2^2 
+ 17328 N g_1^2 g_2^2 \nonumber \\
&& ~~~ + 49536 N g_1 g_2^3 + 663552 \zeta_3 g_2^4 + 1002240 g_2^4 ] 
\frac{1}{32} \nonumber \\
&& +~ [ b_1 g_1^5 + b_2 g_1^4 g_2 + b_3 g_1^3 g_2^2 + b_4 g_1^2 g_2^3
+ b_5 g_1 g_2^4 + b_6 g_2^5] N^3 ~+~ O(g_i^6)
\label{gnybeta}
\end{eqnarray}
where $\zeta_z$ is the Riemann zeta function. We have included two sets of
coefficients $a_i$ and $b_i$ which are the unknown coefficients in each of the 
four loop $\beta$-functions at leading order in large $N$. It may be the case 
that when they are known explicitly that some are actually zero. However 
knowledge of the $O(\epsilon^4)$ term of $\omega_1$ will give a constraint on a
linear combination of the unknown coefficients. To make the connection between 
the $\beta$-functions of (\ref{gnybeta}) and $\omega_1$ is not as 
straightforward as connecting the $\beta$-function of the Gross-Neveu model 
with the analogous large $N$ critical exponent encoding it. This is because in 
that model there is only one coupling constant whereas in the 
Gross-Neveu-Yukawa case there are two coupling constants and hence two 
$\beta$-functions. So in the Gross-Neveu model evaluating the slope of the 
$\beta$-function at criticality it can be compared directly with the $\epsilon$
expansion of the corresponding critical exponent at each order in $1/N$, 
\cite{15,17,18,20}. For a two coupling theory like the Gross-Neveu-Yukawa model
the quantity which is parallel to the $\beta$-function slope is the matrix of 
first order derivatives of $\beta_i(g_1,g_2)$ and defined by  
\begin{equation}
\beta_{ij}(g_1,g_2) ~=~ \left( \frac{\partial \beta_i}{\partial g_j} \right) ~.
\end{equation}
To proceed we find the two eigenvalues of this matrix which are given by 
\begin{eqnarray}
\beta_\pm(g_1,g_2) &=& \frac{1}{2} \left[ \frac{}{} \beta_{11}(g_1,g_2) 
+ \beta_{22}(g_1,g_2) \right. \nonumber \\
&& \left. ~~~ \pm ~ 
\sqrt{ \left[ \left[ \beta_{11}(g_1,g_2) - \beta_{22}(g_1,g_2) \right]^2
+ 4 \beta_{12}(g_1,g_2) \beta_{21}(g_1,g_2) \right] } \right]
\end{eqnarray}
from which we can obtain the two eigen-exponents at criticality defined by
\begin{equation}
\omega_\pm ~=~ \beta_\pm(g_{1c},g_{2c}) 
\end{equation} 
where the subscript ${}_c$ indicates the value of the coupling constants at the
Wilson-Fisher fixed point. The formal large $N$ expansion is given by
\begin{equation}
\omega_\pm ~=~ \sum_{n=0}^\infty \frac{\omega_{\pm \,n}}{N^n} 
\end{equation}
and from (\ref{gnybeta}) we have 
\begin{equation}
\omega_{-\,1} ~=~ -~ 6 \epsilon^2 ~+~ 7 \epsilon^3 ~+~ 
3 \left[ a_1 + \frac{a_2}{2} + \frac{a_3}{4} + \frac{a_4}{8} 
+ \frac{a_5}{16} \right] \epsilon^4 ~+~ O(\epsilon^5) ~.
\end{equation} 
Clearly this is in agreement with $\omega_1$ when allowance is made for the
fact that the $\beta$-functions of (\ref{gnybeta}) were derived for a spinor 
trace convention of $4$. Moreover the $O(\epsilon^4)$ term gives the constraint
\begin{equation} 
a_1 ~+~ \frac{a_2}{2} ~+~ \frac{a_3}{4} ~+~ \frac{a_4}{8} ~+~ 
\frac{a_5}{16} ~=~ \frac{11}{6} 
\end{equation}
which corresponds to a blind test we alluded to earlier. A similar constraint 
can be constructed for the five loop $\beta$-function.

\sect{Non-abelian Nambu-Jona-Lasinio model.}

Having determined $\omega_1$ for the ultraviolet completeness of the 
Gross-Neveu model it is possible to extend the formalism for related $4$-fermi
models with continuous chiral symmetry. These come under the broad title of the
Nambu-Jona-Lasinio model but there are several classes of such models depending
on the group content. Moreover in the more recent language several are the 
ultraviolet completions of two dimensional models which have a different name 
associated with them. For instance the chiral Gross-Neveu (CGN) model is, 
\cite{1,25},
\begin{equation}
L_{\mbox{\footnotesize{CGN}}} ~=~ i \bar{\psi}^i \partialslash \psi^i ~+~ 
g \left[ \sigma \bar{\psi}^i \psi^i ~+~ i \pi \bar{\psi}^i \gamma^5\psi^i
\right] ~-~ \frac{1}{2} [ \sigma^2+\pi^2 ]
\label{lagcgn}
\end{equation}
which is the extension of (\ref{laggn}) from a theory with a discrete chiral
symmetry to one with a continuous $U(1)$ chiral symmetry and involves an
additional bosonic field. The Lagrangian is renormalizable in two dimensions 
and shares other properties with (\ref{laggn}). We refer to it as the chiral
Gross-Neveu model rather than the Nambu-Jona-Lasinio model as the latter is
usually regarded as an effective theory describing meson states in four 
dimensions where it would be non-renormalizable. We will call the ultraviolet 
completion of (\ref{lagcgn}) the chiral Gross-Neveu-Yukawa (CGNY) model which 
has the Lagrangian 
\begin{equation}
L_{\mbox{\footnotesize{CGNY}}} ~=~ 
i \bar{\psi}^i \partialslash \psi^i ~+~ 
\frac{1}{2} \partial_\mu \sigma \partial^\mu \sigma ~+~ 
\frac{1}{2} \partial_\mu \pi \partial^\mu \pi ~+~ 
g_1 \left[ \sigma \bar{\psi}^i \psi^i ~+~ 
i \pi \bar{\psi}^i \gamma^5 \psi^i \right] ~+~ 
\frac{g_2^2}{24} \left( \sigma^2 + \pi^2 \right)^2 
\end{equation}
and has a $U(1)$~$\times$~$U(1)$ chiral symmetry. Like (\ref{laggny})
$L_{\mbox{\footnotesize{CGNY}}}$ has an additional coupling constant compared 
to (\ref{lagcgn}). In the context of extensions which relate more closely to 
the Standard Model the chiral Gross-Neveu model can be endowed with a 
non-abelian symmetry to produce the Lagrangian, \cite{1}, 
\begin{equation}
L_{\mbox{\footnotesize{NJL}}} ~=~ 
i \bar{\psi}^{iI} \partialslash \psi^{iI} ~+~ g \left[ \sigma \bar{\psi}^{iI}
\psi^{iI} ~+~ i \pi^a \bar{\psi}^{iI} \lambda^a_{IJ}\gamma^5 \psi^{iJ} 
\right] ~-~ \frac{1}{2} \left[ \sigma^2 + {\pi^a}^2 \right]
\label{lagnjl}
\end{equation}
where $T^a$~$=$~$\half \lambda^a$ are the Lie group generators. The additional 
indices have ranges $1$~$\leq$~$I$~$\leq$ $M$ and $1$~$\leq$~$a$~$\leq$~$\NA$ 
where $\NA$ is the dimension of the adjoint representation of the Lie group. We
will call this the non-abelian Nambu-Jona-Lasinio model or universality class. 
This version was introduced in \cite{1} where it was shown to be invariant
under chiral $SU(M)$~$\times$~$SU(M)$ transformations. We use this form, 
(\ref{lagnjl}), in order to be consistent with the notation of earlier large 
$N$ computations of critical exponents. The conventions of \cite{26,27} are
retained and we recall that the Lie group Casimirs are
\begin{equation}
\mbox{Tr} (\lambda^a \lambda^b) ~=~ 4T_F \delta^{ab} ~~~,~~~
\lambda^a \lambda^a ~=~ 4 C_F I ~~~,~~~ f^{acd} f^{bcd} ~=~ C_A \delta^{ab} ~.
\end{equation}
Then the ultraviolet completeness of (\ref{lagnjl}) is the non-abelian
Nambu-Jona-Lasinio-Yukawa model
\begin{eqnarray}
L_{\mbox{\footnotesize{NJLY}}} &=&
i \bar{\psi}^{iI} \partialslash \psi^{iI} ~+~ 
\frac{1}{2} \partial_\mu \sigma \partial^\mu \sigma ~+~ 
\frac{1}{2} \partial_\mu \pi^a \partial^\mu \pi^a \nonumber \\
&& +~ g_1 \left[ \sigma \bar{\psi}^{iI} \psi^{iI} ~+~ 
i \pi^a \bar{\psi}^{iI} \lambda^a_{IJ}\gamma^5 \psi^{iJ} \right] ~+~ 
\frac{g_2^2}{24} \left( \sigma^2 + {\pi^a}^2 \right)^2 ~.
\end{eqnarray}
Given the similarity of the Lagrangians (\ref{lagcgn}) and (\ref{lagnjl}) we 
will focus on computing $\omega_1$ in the latter because the chiral Gross-Neveu
Lagrangian follows by taking the abelian limit of the non-abelian 
Nambu-Jona-Lasinio model. By this we mean that formally $\lambda^a$ is replaced
by the unit matrix and $\pi^a$ by the $\pi$ field in (\ref{lagnjl}). For the 
resulting exponents this will correspond to the Casimirs taking the values 
$C_F$~$=$~$T_F$~$=$~$\quarter$, $C_A$~$=$~$0$ and $M$~$=$~$1$. 

As the large $N$ method for the non-abelian Nambu-Jona-Lasinio universal theory
at the Wilson-Fisher fixed point is analogous to that used for the Gross-Neveu 
case we will indicate the major differences which have to be taken into account
in order to deduce $\omega_1$. First, as there are three fields now the 
asymptotic scaling forms for the propagators and $2$-point functions are,
\cite{26,27},
\begin{eqnarray}
\psi(x) &\sim& \frac{A\xslash}{(x^2)^\alpha} \left[ 1 + A^\prime(x^2)^\omega
\right] ~~~,~~~
\sigma(x) ~\sim~ \frac{B}{(x^2)^\beta} \left[ 1 + B^\prime(x^2)^\omega
\right] \nonumber \\
\pi(x) &\sim& \frac{C}{(x^2)^\gamma} \left[ 1 + C^\prime(x^2)^\omega \right] 
\end{eqnarray}
and
\begin{eqnarray}
\psi^{-1}(x) & \sim & \frac{r(\alpha-1)\xslash}{A(x^2)^{2\mu-\alpha+1}}
\left[ 1 - A^\prime s(\alpha-1)(x^2)^\omega \right] \\
\sigma^{-1}(x) & \sim &\frac{p(\beta)}{B(x^2)^{2\mu-\beta}}
\left[ 1 - B^\prime q(\beta) (x^2)^\omega \right] \nonumber \\
\pi^{-1}(x) & \sim &\frac{p(\gamma)}{C(x^2)^{2\mu-\gamma}}
\left[ 1 - C^\prime q(\gamma) (x^2)^\omega \right]
\end{eqnarray}
respectively in coordinate space where $C$ and $C^\prime$ are additional 
$x$-independent amplitudes. In this section our notation will primarily follow 
that of \cite{26,27}. The amplitude $C$ will appear within the skeleton
Schwinger-Dyson equations in the combination $y$~$=$~$A^2C$ analogous to $z$.
The dimensions of the respective fields are defined as  
\begin{equation}
\alpha ~=~ \mu ~+~ \half \eta ~~~,~~~ 
\beta ~=~ 1 ~-~ \eta ~-~ \chi_\sigma ~~~,~~~ \gamma ~=~ 1 ~-~ \eta ~-~ \chi_\pi
\end{equation}
where $\chi_\pi$ is the anomalous dimension of the $3$-point vertex involving
$\pi^a$. With the additional $\pi^a$ field the set of skeleton Schwinger-Dyson
equation changes and these are illustrated in Figure $5$. The various two loop
graphs in the $\sigma$ and $\pi^a$ $2$-point functions are again needed due to
the same reordering as the Gross-Neveu model which is present here too. 

{\begin{figure}[ht]
\begin{center}
\includegraphics[width=15cm,height=6cm]{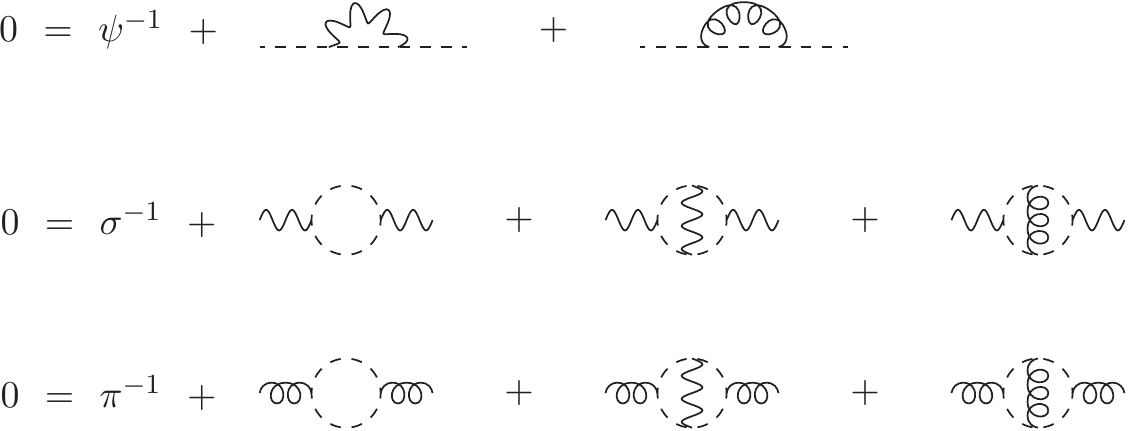}
\end{center}
%\vspace{0.5cm}
\caption{Leading order skeleton Schwinger-Dyson equations used to determine
$\omega_1$ in the non-abelian Nambu-Jona-Lasinio model.}
\end{figure}}

There are no mixed $\sigma$-$\pi^a$ $2$-point functions as would be allowed by 
the Feynman rules. This is because of the preservation of the chiral symmetry
and general properties of $\gamma^5$. For instance, such mixing terms could be 
chirally rotated back to the set of Figure $5$. Moreover the role and presence 
of $\gamma^5$ in the $d$-dimensional context deserves comment. While there is 
no concept of chiral symmetry beyond even integer dimensions the matrix 
$\gamma^5$ is retained but is regarded as a formal object which obeys the naive 
anti-commutation relation
\begin{equation}
\{ \gamma^\mu , \gamma^5 \} ~=~ 0
\end{equation}
in $d$-dimensions when an even number of $\gamma^5$'s are present. The other 
important aspect to stress is that while we are performing computations in 
$d$-dimensions the spacetime dimension has not been extended away from its
critical dimension in order to effect a regularization of the theory in
perturbation theory. In the large $N$ expansion method developed in 
\cite{12,13} divergent Feynman integrals are regularized analytically by 
shifting the vertex anomalous dimensions $\chi_i$~$\to$~$\chi_i$~$+$~$\Delta$ 
where $\Delta$ is the regularizing parameter. While the integrals we need here 
are finite in the large $N$ context, and hence do not need to be regularized, 
we mention the large $N$ regularization since it obviates the need to discuss 
how $\gamma^5$ is defined in this formalism. It is regarded as an object which 
obeys the above simple algebra in $d$-dimensions. This is consistent, for 
example, with the use of the $\epsilon$ expansion of critical exponents. Such 
expansions are invariably derived from theories in either two or four 
dimensions and then the $\epsilon$ expansion is summed before deducing an 
estimate in three dimensions. Strictly when one constructs such an $\epsilon$ 
expansion one has lost the chiral symmetry of the fixed even dimensional 
spacetime and then it is not the chiral symmetry as such which is the main 
property. Rather it is the preservation of the underlying anti-commutativity of
$\gamma^5$ to bridge across the dimensions between two and four, and its 
properties to exclude a mixed $\sigma$-$\pi^a$ $2$-point function for example, 
which is the main guiding principal in $d$-dimensions. This anti-commutativity 
was also central to the associated large $N$ computations here.

Like the Gross-Neveu universality class we need the basic quantities for 
(\ref{lagnjl}) in the large $N$ expansion. The leading order diagrams of 
Figure $5$ with the first terms of the scaling functions give, \cite{26,27}, 
\begin{equation}
0 ~=~ r(\alpha-1) ~+~ z ~+~ 4 C_F y ~~~,~~~
0 ~=~ p(\beta) ~+~ 2 M N z ~~~,~~~
0 ~=~ p(\gamma) ~+~ 8 T_F N y
\end{equation}
from which it follows that, \cite{26,27}, 
\begin{equation}
\eta_1 ~=~ \frac{\tilde{\eta}_1}{2} \left[ \frac{1}{M} ~+~ \frac{C_F}{T_F} 
\right] ~.
\end{equation}
We have introduced the group independent exponent 
\begin{equation}
\tilde{\eta}_1 ~=~ -~ \frac{2\Gamma(2\mu-1)}{\Gamma(\mu+1)\Gamma(\mu)
\Gamma(1-\mu)\Gamma(\mu-1)}
\end{equation}
which is core to both the chiral Gross-Neveu and non-abelian Nambu-Jona-Lasinio 
classes. Equally 
\begin{equation}
z_1 ~=~ \frac{\mu\Gamma^2(\mu)\tilde{\eta}_1}{4M} ~~~,~~~
y_1 ~=~ \frac{Mz_1}{4T_F}
\end{equation}
and 
\begin{equation}
\chi_{\sigma \, 1} ~=~ \frac{\mu\tilde{\eta}_1}{2(\mu-1)} 
\left[ \frac{1}{M} - \frac{C_F}{T_F} \right] ~~~,~~~
\chi_{\pi \, 1} ~=~ \frac{\mu\tilde{\eta}_1}{2(\mu-1)} \left[ \frac{C_F}{T_F}
- \frac{1}{M} - \frac{C_A}{2T_F} \right] 
\end{equation}
which are required to determine $\omega_1$. For orientation and for comparison
to the Gross-Neveu skeleton Schwinger-Dyson equations at criticality we give
the algebraic representation of the first two equations of Figure $5$ which
are, \cite{26,27}, 
\begin{eqnarray}
0 &=& r(\alpha-1) \left[ 1-A^\prime s(\alpha-1)(x^2)^\omega \right] ~+~ 
z \left[ 1+(A^\prime + B^\prime)(x^2)^\omega \right] \nonumber \\
&& +~ 4 C_F y  \left[ 1+(A^\prime + C^\prime)(x^2)^\omega \right] \nonumber \\
0 &=& \frac{p(\beta)}{N} \left[ 1-B^\prime q(\beta)(x^2)^\omega \right] ~+~ 
2zM \left[ 1 + 2A^\prime(x^2)^\omega  \right] ~-~ 
z^2M \left[ \Pi + ( A^\prime\Pi_{1A}+B^\prime\Pi_{1B} ) (x^2)^\omega \right] 
\nonumber \\
&& +~ 4yzM C_F \left[ \Pi + (\Pi_{1A} A^\prime ~+~ 
\Pi_{1C} C^\prime )(x^2)^\omega \right]
\end{eqnarray}
where $\Pi_{1C}$ is the value of the analogous graph to that illustrated in
Figure $2$. Although the leading order value in the $1/N$ expansion is the same
as $\Pi_{1B}$ the corrections will be different since the value of the vertex 
anomalous dimension are different for the $3$-point vertices. With the extra 
field present the consistency equation to deduce $\omega_1$ is derived from 
ensuring that the determinant of the matrix formed by the three equations with 
respect to the vector $(A^\prime, B^\prime, C^\prime)$ is zero. The three
equations are 
\begin{eqnarray}
0 &=& \left[ z + 4 C_F y - r(\alpha-1)s(\alpha-1) \right] A^\prime ~+~ 
z B^\prime ~+~ 4 C_F y C^\prime \nonumber \\
0 &=& 4 z A^\prime ~-~ \left[\frac{p(\beta)q(\beta)}{NM}
+ z^2 \Pi_{1B}\right] B^\prime ~+~ 4 C_F yz\Pi_{1C} C^\prime \nonumber \\
0 &=& 16 T_F y A^\prime ~+~ 4 T_F yz\Lambda_{1B}B^\prime ~-~ 
\left[ \frac{p(\gamma)q(\gamma)}{N} ~+~ 8 [ 2C_F - C_A ] T_F y^2 \Lambda_{1C} 
\right] C^\prime
\end{eqnarray}
where we designate the two loop correction graphs analagous to that of Figure
$2$ but with $\pi^a$ external legs by the letter $\Lambda$, \cite{27}. Again 
the leading order large $N$ values of $\Lambda_{1B}$ and $\Lambda_{1C}$ are the 
same as $\Pi_{1B}$. Solving the determinant at leading order produces  
\begin{equation}
r(\alpha-1) s(\alpha-1) \left[ 2 q(\beta) ~-~ z \left[ 1 - \frac{C_F M}{T_F}
\right] \Pi_{1B} \right] ~=~ 
4 z \left[ 1 + \frac{C_FM}{T_F} \right]
\end{equation}
which contains $\omega_1$ as the only unknown where we have rewritten all the 
two loop graphs in terms of $\Pi_{1B}$, with value given in (\ref{valpi1b}),
which is valid at this order in the $1/N$ expansion. This would not be valid 
for finding $\omega_2$. Finally, this produces the result  
\begin{equation}
\omega^{\mbox{\footnotesize{NJL}}}_1 ~=~ -~ \frac{(2\mu-1)(\mu-2)}{(\mu-1)} 
\left[ \frac{1}{M} + (\mu-2) \frac{C_F}{T_F} \right] \tilde{\eta_1} ~.
\end{equation} 
Taking the abelian limit gives the analogous quantity for (\ref{lagcgn}) which
is
\begin{equation}
\omega^{\mbox{\footnotesize{CGN}}}_1 ~=~ -~ (2\mu-1)(\mu-2) \tilde{\eta_1} ~.
\end{equation} 
From these we find
\begin{eqnarray}
\omega^{\mbox{\footnotesize{CGN}}}_1 &=& 12 \epsilon^2 ~-~ 26 \epsilon^3 ~+~
3 \epsilon^4 ~+~ O(\epsilon^5) \nonumber \\ 
\omega^{\mbox{\footnotesize{NJL}}}_1 &=& \frac{12}{M} \epsilon^2 ~-~ 
\left[ 12 \frac{C_F}{T_F} + \frac{14}{M} \right] \epsilon^3 ~+~
\left[ 14 \frac{C_F}{T_F} - \frac{11}{M} \right] \epsilon^4 ~+~ O(\epsilon^5)
\end{eqnarray} 
when $d$~$=$~$4$~$-$~$2\epsilon$. Equally  
\begin{eqnarray}
\left. \frac{}{} \omega^{\mbox{\footnotesize{CGN}}} \right|_{d=3} &=& 
-~ \frac{1}{2} ~+~ \frac{16}{3\pi^2 N} ~+~ O \left( \frac{1}{N^2} \right) 
\nonumber \\
\left. \frac{}{} \omega^{\mbox{\footnotesize{CGN}}} \right|_{d=5} &=& 
\frac{1}{2} ~+~ \frac{256}{5\pi^2 N} ~+~ O \left( \frac{1}{N^2} \right) 
\end{eqnarray}
and
\begin{eqnarray}
\left. \frac{}{} \omega^{\mbox{\footnotesize{NJL}}} \right|_{d=3} &=& 
-~ \frac{1}{2} ~-~ \frac{16[C_FM-2T_F]}{3\pi^2 T_F M N} ~+~ 
O \left( \frac{1}{N^2} \right) \nonumber \\
\left. \frac{}{} \omega^{\mbox{\footnotesize{NJL}}} \right|_{d=5} &=& 
\frac{1}{2} ~+~ \frac{256[2T_F+C_FM]}{15\pi^2 T_F M N} ~+~ 
O \left( \frac{1}{N^2} \right) 
\end{eqnarray}
for various odd dimensions. 

\sect{Discussion.} 

We have evaluated the critical exponent $\omega$ at $O(1/N)$ in the Gross-Neveu 
universality class as a function of $d$. It is the exponent which relates to 
the $\beta$-functions of the four dimensional Gross-Neveu-Yukawa theory which 
is not unrelated to the Standard Model. In addition we have found $\omega_1$ 
for two other cases which are the chiral Gross-Neveu and non-abelian 
Nambu-Jona-Lasinio universality classes. The actual large $N$ computation 
differed from the parallel one for the $\beta$-function exponent of the two 
dimensional Gross-Neveu model itself in that due to a reordering within the 
critical point large $N$ formalism a {\em two} loop graph had to be determined 
in order to deduce $\omega_1$. While it was a straightforward exercise to 
achieve this using conformal integration methods for massless integrals it 
means that the evaluation of $\omega_2$ will involve the computation of various
three and four loop Feynman graphs contributing to the $\sigma$ skeleton 
Schwinger-Dyson equation. This would be the natural extension of the present
work. Equally the comparison with four dimensional perturbation theory was not 
as straightforward compared to the two dimensional case. The Gross-Neveu-Yukawa
model has two coupling constants and hence two $\beta$-functions which means 
that the comparison of the large $N$ exponent is with one of the eigenvalues of
the Hessian of the $\beta$-functions at criticality. In terms of the underlying
universal theory the two coupling constants can be viewed in a different way. 
Clearly the Gross-Neveu-Yukawa theory is renormalizable in four dimensions. 
However, away from four dimensions the operators associated with each coupling 
constant are different in $d$-dimensions and hence the degeneracy is lifted. At
the Wilson-Fisher fixed point, which the large $N$ formalism used here operates
at, the core interaction which drives the Gross-Neveu universality class is the
$\sigma \bar{\psi} \psi$ one. The quartic $\sigma$ interaction in some sense 
spectates in four dimensions but is either relevant or irrelevant away from 
this dimension. Another way of putting this is that the canonical dimensions of
each operator are the same in four dimensions but away from four dimensions 
they are no longer equal and therefore there is no degeneracy of the couplings.
In other words in the dimension where the operators overlap there is a mixing 
which manifests itself as two $\beta$-functions in perturbation theory but 
operator mixing in the large $N$ universal theory. This is the reason why we 
had the more involved comparison with perturbation theory in contrast to 
\cite{15,17,18,20}. Viewing this aspect of the Gross-Neveu-Yukawa theory it is 
not inconsistent with the emergent supersymmetry which has been noticed 
recently, \cite{7}. It is well known that supersymmetry is a symmetry of an 
integer dimension. Away from say four dimensions the degrees of freedom of the 
bosons and fermions are no longer equivalent. Indeed to consistently 
dimensionally regularize a supersymmetric theory is not possible because of 
this. One method to ameliorate this loss of equality was devised in \cite{28} 
and requires extra fields which live only in the dimensionally extended part of
the spacetime. At the emergent supersymmetric fixed point one would have to 
preserve the equivalence along the Wilson-Fisher fixed point in $d$-dimensions 
of its corresponding universal theory.

\vspace{1cm}
\noindent
{\bf Acknowledgements.} The author thanks R.M. Simms, Dr L. Mihaila and Dr M. 
Scherer for valuable discussions. The work was carried out with the support of 
the STFC through the Consolidated Grant ST/L000431/1. The graphs were drawn 
with the {\sc Axodraw} package \cite{29}.

\end{document}